\documentclass[%
 aip,
 amsmath,amssymb,
 reprint,%
]{revtex4-1}

\usepackage{graphicx}
\usepackage{dcolumn}
\usepackage{bm}

\usepackage[utf8]{inputenc}
\usepackage[T1]{fontenc}
\usepackage{mathptmx}
\usepackage{etoolbox}
\usepackage{float}
\usepackage{natbib}
\usepackage{hyperref}

\makeatletter
\def\@email#1#2{%
 \endgroup
 \patchcmd{\titleblock@produce}
  {\frontmatter@RRAPformat}
  {\frontmatter@RRAPformat{\produce@RRAP{*#1\href{mailto:#2}{#2}}}\frontmatter@RRAPformat}
  {}{}
}%
\makeatother

\begin{document}

\preprint{AIP/123-QED}

\title[Draft]{Minimizing the Ground Effect for Photophoretically Levitating Disks}
\author{Zhipeng Lu}
 \affiliation{Department of Chemistry, University of Pennsylvania.}
 \affiliation{ 
Department of Mechanical Engineering and Applied Mechanics, University of Pennsylvania.
}%
 \email{zplu@seas.upenn.edu.}

\author{Miranda Stern}%
\affiliation{ 
Department of Mechanical Engineering and Applied Mechanics, University of Pennsylvania.
}%

\author{Jinqiao Li}%
\affiliation{ 
Department of Materials Science and Engineering, University of Pennsylvania.
}%

\author{David Candia}%
\affiliation{ 
Department of Mechanical Engineering and Applied Mechanics, University of Pennsylvania.
}%

\author{Lorenzo Yao-Bate}%
\affiliation{ 
Department of Mechanical Engineering and Applied Mechanics, University of Pennsylvania.
}%
\affiliation{ 
Department of Physics, University of Pennsylvania.
}%

\author{Thomas J. Celenza}%
\affiliation{ 
Department of Mechanical Engineering and Applied Mechanics, University of Pennsylvania.
}%

\author{Mohsen Azadi}%
\affiliation{ 
Singh Center for Nanotechnology, University of Pennsylvania.
}%

\author{Matthew F. Campbell}%
\affiliation{ 
Department of Mechanical Engineering and Applied Mechanics, University of Pennsylvania.
}%

\author{Igor Bargatin}%
\affiliation{ 
Department of Mechanical Engineering and Applied Mechanics, University of Pennsylvania.
}%

\date{\today}

\begin{abstract}
Photophoretic levitation is a propulsion mechanism in which lightweight objects can be lifted and controlled through their interactions with light. Since photophoretic forces on macroscopic objects are usually maximized at low pressures, they may be tested in vacuum chambers in close proximity to the chamber floor and walls. We report here experimental evidence that the terrain under levitating microflyers, including the chamber floor or the launchpad from which microflyers lift off, can greatly increase the photophoretic lift forces relative to their free-space (mid-air) values. To characterize this so-called “ground effect” during vacuum chamber tests, we introduced a new miniature launchpad composed of three J-shaped (candy-cane-like) wires that minimized a microflyer’s extraneous interactions with underlying surfaces. We compared our new launchpads to previously used wire-mesh launchpads for simple levitating mylar-based disks with diameters of 2, 4, and 8 cm. Importantly, wire-mesh launchpads increased the photophoretic lift force by up to sixfold. A significant ground effect was also associated with the bottom of the vacuum chamber, particularly when the distance to the bottom surface was less than the diameter of the levitating disk. We provide guidelines to minimize the ground effect in vacuum chamber experiments, which are necessary to test photophoretic microflyers intended for high-altitude exploration and surveillance on Earth or on Mars.
\end{abstract}

\maketitle

\section{\label{sec:level1}Introduction}
Microflyers are typically defined as airborne vehicles with dimensions smaller than ~10 cm. Compared to conventional unmanned aerial vehicles (UAVs) like weather balloons, drones, and satellites, microflyers are lightweight, low-cost, and use less energy. Some microflyers can be driven by wind or solar energy and may not need a battery, engine, or motor. Such miniaturized unpowered aerial vehicles can be dispersed by wind as plant seed-inspired microrobots \cite{three-dimensional} or powered by the Sun using photovoltaics \cite{untethered-flight} or photophoretic levitation \cite{controlled-levitation}, potentially enabling applications in ubiquitous sensing and monitoring of the atmosphere, climate, and local environment \cite{unmanned-aerial-systems, unmanned-aerial-vehicle, an-overview}. 

Although microflyers are typically designed for uses far from the ground, they are often tested in close proximity to horizontal surfaces, whether solid or very sparse like a wire mesh launchpad \cite{photophoretic-levitation, controlled-levitation, influence-of}. These underlying surfaces can greatly modify the airflow in the vicinity of the microflyers, create an area of increased pressure under the microflyers, and enhance the lift force experienced by the microflyers in a phenomenon called the ground effect \cite{wing-in, design-of, aerodynamic-investigation}. In the continuum regime, i.e., when the mean free path is much smaller than the flyer dimensions, the ground effect has been well studied for three-dimensional hovercraft and aircraft that take off or cruise at very low altitudes. Previous researchers mainly focused on how the enhanced lift can be used to address flight security, fuel consumption \cite{wing-in, ground-effect}, and control of mini-rotorcraft \cite{mitigating-ground}. 

The ground effect can also be significant in photophoretic levitation, as previously calculated for the photophoresis of microscopic aerosol spheres close to a solid plane surface \cite{photophoresis-of} and observed for macroscopic plates hovering on an air cushion at atmospheric pressure \cite{photophoretic-levitation}. During testing of photophoretic microflyers, the mean free path is often comparable to the characteristic dimension of the flyer, i.e., the experiments are done in the transition regime between free-molecular and continuum fluid dynamics. The ground effect, therefore, results from a combination of free-molecular back-and-forth bouncing of air molecules between the flyer and the launchpad and continuum air flow (e.g., an elevated-pressure air cushion) \cite{photophoretic-levitation, photophoresis-of}. The associated increase of the lift force can be particularly large when tests are done in a typical tabletop vacuum chamber in which the distance to the nearest horizontal surface may be comparable to or even smaller than the levitating vehicle’s characteristic size. For macroscopic photophoretic flyers, the aerodynamic differences between laboratory test launchpads and real-world mid-air environment may therefore lead to exaggerated expectations for microflyers’ altitude range and payload capability. 

We are unaware of any prior work characterizing the ground effect in the transition regime for ultralight microflyers, whether due to launchpads or other chamber-related terrain. Given the critical importance of accurate tests for future microflyer deployment on Earth or Mars, in this report we quantify the impact of the ground effect for both traditional wire mesh-based launchpads and for a new minimal launchpad that we describe below. 

\section{\label{sec:level1}Results and Discussions}
We conducted the experiments in this work using simple microflyers consisting of disk-shaped 0.5-micron-thick mylar films with diameters of 2, 4, and 8 cm. These microflyers were chosen due to their ease of fabrication and adequate mechanical, thermal, and aerodynamic stability \cite{controlled-levitation}. Our levitation mechanism relies on a difference in the thermal accommodation coefficient (TAC) across these films, made possible by depositing a thin carbon nanotube (CNT) film on the underside of the disks \cite{controlled-levitation, semi-emipirical, photophoretic-force, photophoresis-on, re-examination, theory-of, measurement-of, method-for}. A representative 8-cm-diameter disk is shown in Fig. 1(d). We test our microflyers in an acrylic vacuum chamber positioned over an array of eight light emitting diodes (LEDs) as shown in Fig. 1(a) and Fig. S1(a, b). 

Our primary experimental variables are the light irradiance experienced by the microflyers and the vacuum chamber pressure. As discussed in the Supplemental Material, we carefully characterized the light irradiance as a function of position, elevation, and any shading factors present (such as the launchpad) within the vacuum chamber using a photodiode and an optical sensor. When illuminated by incident light, the microflyer experiences (1) an upward photophoretic force, which may be enhanced by the ground effect, (2) a downward gravitational pull, and (3) an electrostatic stiction force that can be either repulsive (upward) or attractive (downward). The photophoretic force increases with the light irradiance and is typically maximized at pressures where the mean free path is comparable to the disk diameter (i.e., for 0.01 < Kn < 10, where Kn is the Knudsen number) \cite{photophoresis-of}. This implies that, for a given microflyer size, a pressure exists at which the irradiance needed for liftoff is minimized. A single experiment involves increasing the light irradiance until the microflyer lifts off while holding the chamber pressure constant. This process can be repeated at different pressures to determine the pressure that minimizes the irradiance required for lift off, or the so-called \textit{optimal pressure} (see, e.g., Fig. 2). After measuring this minimum light irradiance as a function of pressure for a variety of microflyers and launchpads, we can infer the magnitude of the ground effect for each microflyer-launchpad combination from its optimal pressure and corresponding light irradiance. For instance, consider the case in which the minimum light irradiance required for liftoff at a given pressure decreases when the launchpad is altered. Given the approximate linear relationship between light irradiance and the photophoretic lift force in the transition regime \cite{semi-emipirical, controlled-levitation}, we can infer that the ground effect is enhancing the lift force more when using the altered launchpad compared to the unmodified launchpad. More details will be provided later in the paper. 

Achieving liftoff in levitation experiments is facilitated by launchpads that elevate microflyers and minimize electrostatic stiction forces associated with the chamber floor. Although not shown in the figures, we connected every launchpad to an electric grounding circuit to minimize these stiction forces. Previously used first-generation launchpads \cite{controlled-levitation, photophoretic-levitation} are composed of sparse steel mesh grids that allow a very high fraction of the incident light to pass through. The main geometric parameters of the mesh are wire diameter $d$ and wire spacing $s$, resulting in open area fraction $\Phi =s^2/(s+d)^2$ for a square mesh, as shown in Fig. 1(b). Note that the quantity $1-\Phi$ represents both the shading of the microflyer by the mesh and the area most closely associated with the ground effect.  

\begin{figure*}[hbt!]
\centering
\includegraphics[width=\textwidth]{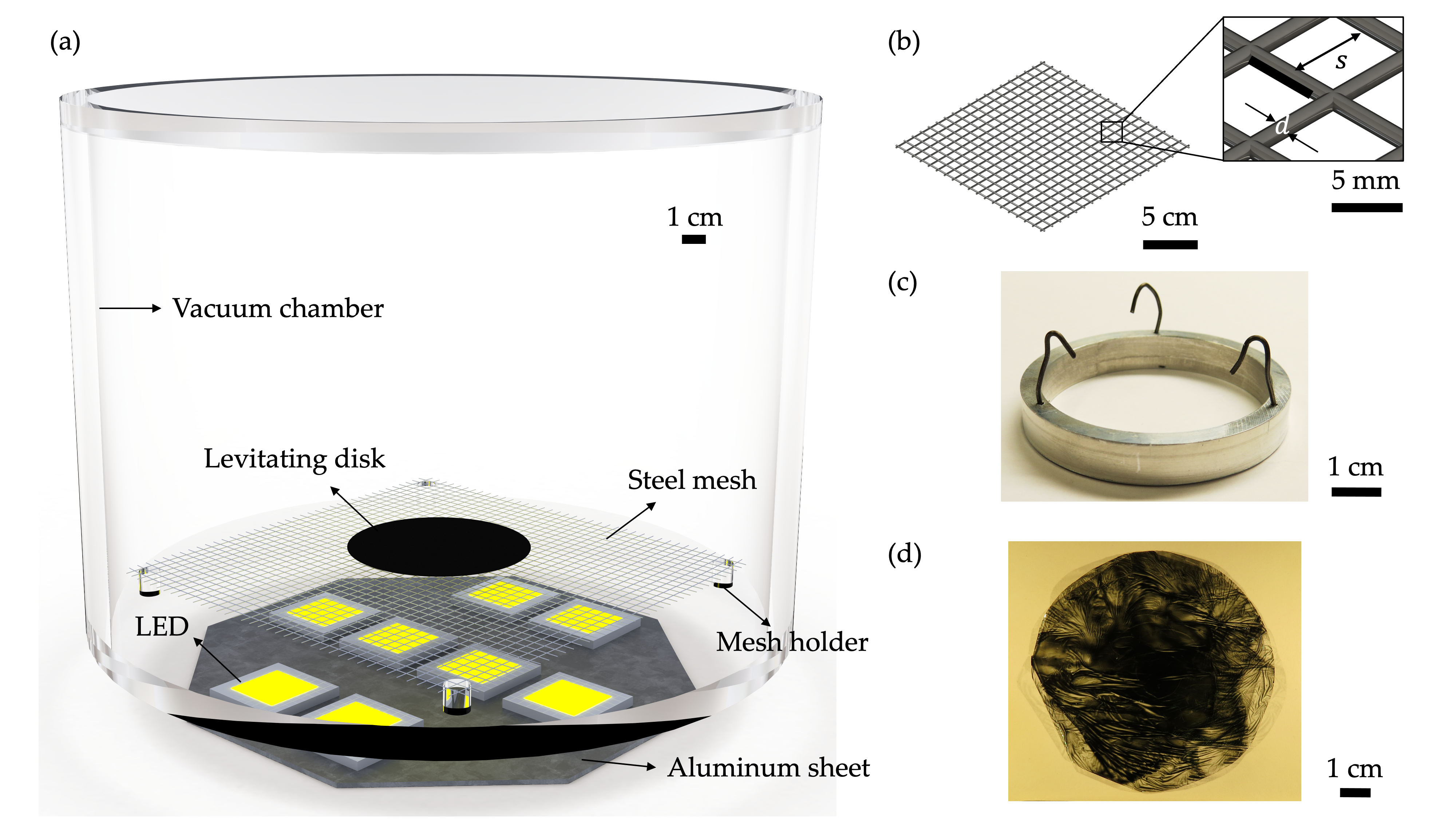}
\caption{Experimental setups of first and second-generation launchpads and a microflyer. (a) Schematic diagram of the experimental setup consisting of an acrylic vacuum chamber, an 8-cm-diameter CNT-mylar-alumina microflyer, a piece of 73\%-open steel mesh, and eight LEDs below the chamber. (b) Schematic diagram of a 0.9-mm-diameter 5.4-mm-spacing steel mesh showing wire spacing $s$ and diameter $d$. (c) Photograph of a triad of 0.9-mm-diameter J-shaped steel sticks and the holder ring. (d) Photograph of an 8-cm-diameter CNT-mylar-alumina disk. Electric grounding circuits are not shown for both types of launchpads for simplicity. }
\end{figure*}

We first examined wire-mesh launchpads, previously used in Refs. [3] and [7], by systematically varying the wire diameter and spacing.  Specifically, we used three steel meshes with the same 0.9-mm wire diameter and wire spacings of 1.0, 3.5, and 5.4 mm (with open area percentages for these spacings of 28\%, 63\%, and 73\%, respectively). As seen from Fig. 2, the central values of optimal pressures for all three spacings fell between 10-14 Pa, with no obvious dependence on the wire spacing and microflyer disk diameter. Notice that the optimal pressures in Fig. 2 are marked as dark-shaded intervals due to the uncertainty produced by experimental observation error and the locally estimated scatterplot smoothing (LOESS) algorithm we used. Without ground effect, i.e., in mid-air, larger samples are predicted to have proportionally lower optimal pressures than smaller samples due to their larger characteristic lengths \cite{photophoresis-of, semi-emipirical}. The fact that we do not observe this in Fig. 2 indicates that ground effects are impacting the photophoretic lift force. The minimum light irradiance also increased as the steel mesh became more open, which is expected because the launchpad-associated ground effect should vanish as the launchpad tends to 100\% open area. We will discuss the physical mechanism of the ground effect later in this paper.

\begin{figure*}[hbt!]
\centering
\includegraphics[width=\textwidth]{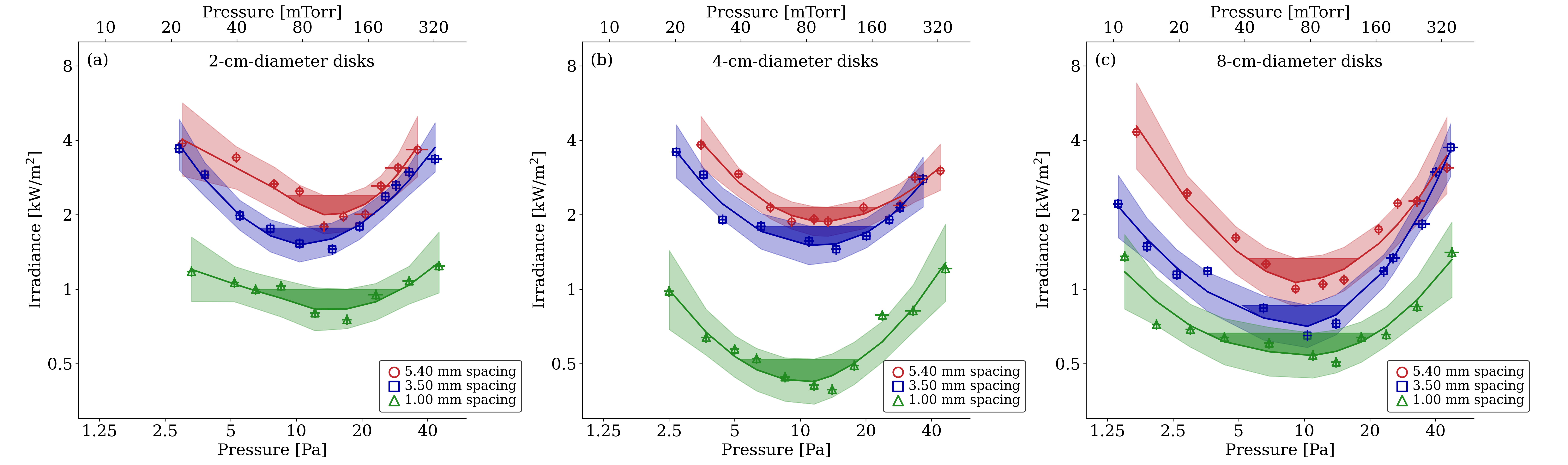}
\caption{Levitation performance of microflyers on steel meshes with a fixed wire diameter (0.9 mm) but changing wire spacings. The optimal pressure can be seen to be independent of the wire spacing and the microflyer diameter. Microflyers are (a) 2, (b) 4, and (c) 8-cm-diameter alumina-mylar-CNT disks. Solid lines stand for LOESS fits, with 99\% confidence intervals shown with light shading. Optimal pressures are shown with darker shading.}
\end{figure*}

Next, we fixed the wire spacing to be 3.5 mm and varied the wire diameter instead. The open area percentages for the 0.23, 0.9, and 1.6-mm-diameter wires that we selected were 88\%, 63\%, and 47\%, respectively. As shown in Fig. 3, the optimal pressure increased by roughly a factor of two, in agreement with the roughly twofold decrease in the open area as the wire diameter decreased (by a factor of seven). In addition, the minimum irradiance for lift-off increased as the wire diameter decreased (i.e., as the open area increased). The decreasing influence of the ground effect with increasing open area suggests that ideal launchpads be very sparse. Moreover, some of the experimental results that we have previously published using mesh launchpads may be subject to ground effect lift enhancements \cite{controlled-levitation, photophoretic-levitation}, as discussed in the Supplemental Material.

\begin{figure*}[hbt!]
\centering
\includegraphics[width=\textwidth]{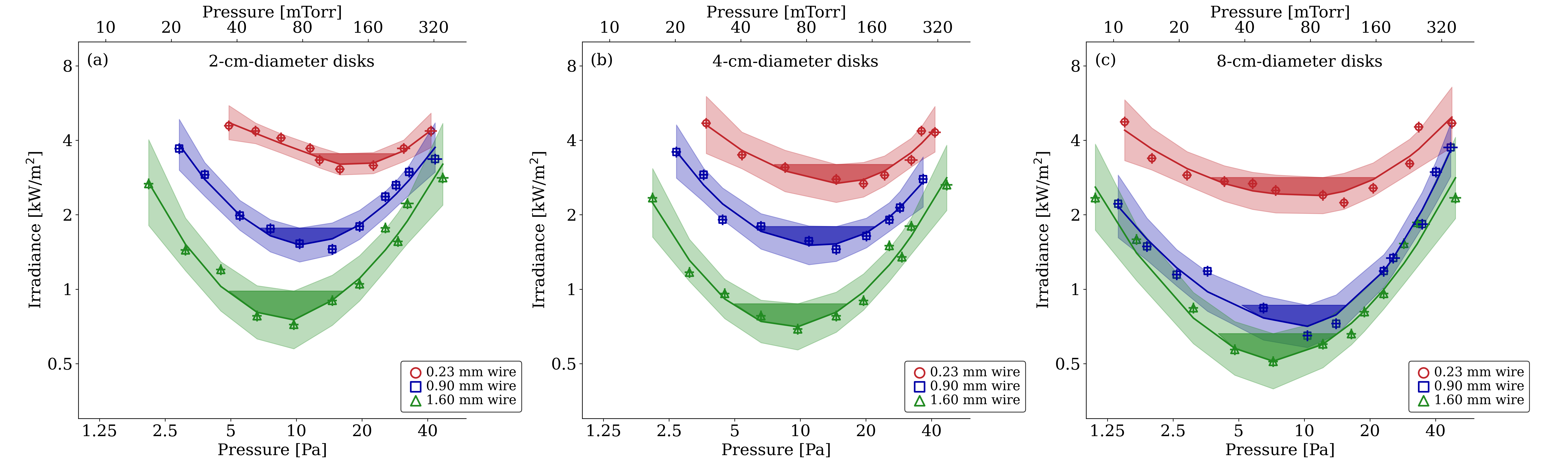}
\caption{Same as Fig. 2 for steel meshes with a fixed wire spacing (3.5 mm) but changing wire diameters. The optimal pressure decreases with increasing wire diameter (i.e., decreasing open area).}
\end{figure*}

To mitigate the ground effect in future experiments, we designed a second-generation launchpad consisting of three J-shaped (i.e., candy-cane-like) wires that minimized the contact area with the microflyer, as shown in Fig. 1(c). We used wires from the same meshes as in Fig. 3 (i.e., with diameters of 0.23, 0.9, and 1.6 mm), bent them into “J” shapes with a millimeter-scale radius of curvature at the top, and carefully maintained the same chamber floor-to disk distance as the mesh launchpads used. A triad of wire canes inserted vertically into an aluminum ring holder provided the necessary stability to support microflyers before takeoff while minimizing the interfacial contact. The effective open area was greater than 99\% even for the thickest 1.6-mm-diameter wire and the smallest 2-cm-diameter disk. 

Our experiments indicate that the second-generation launchpads showed minimal ground effect due to the underlying J-shaped wires. In contrast to the wire mesh launchpads, we observed no obvious dependence on wire diameter for J-shaped-wire launchpads made from the 0.23, 0.9, and 1.6-mm-diameter wires (Fig. 4). Instead, the optimal pressure now showed a clear dependence on the diameter of the microflyer itself, as predicted by the theory of mid-air levitation \cite{controlled-levitation}. Average optimal pressures for 2, 4, and 8-cm-diameter disks centered around 9.6, 6.1, and 3.5 Pa, respectively. The corresponding mean free paths at room temperature for these pressures are approximately 0.73, 1.2, and 2.0 mm, resulting in optimal Knudsen numbers of 0.036, 0.030, and 0.025, respectively. It is well known that the mid-air photophoretic force is maximized in transition regime \cite{photophoresis-of}, or Knudsen numbers from 0.01 to 10, in agreement with our findings. 

\begin{figure*}[hbt!]
\centering
\includegraphics[width=\textwidth]{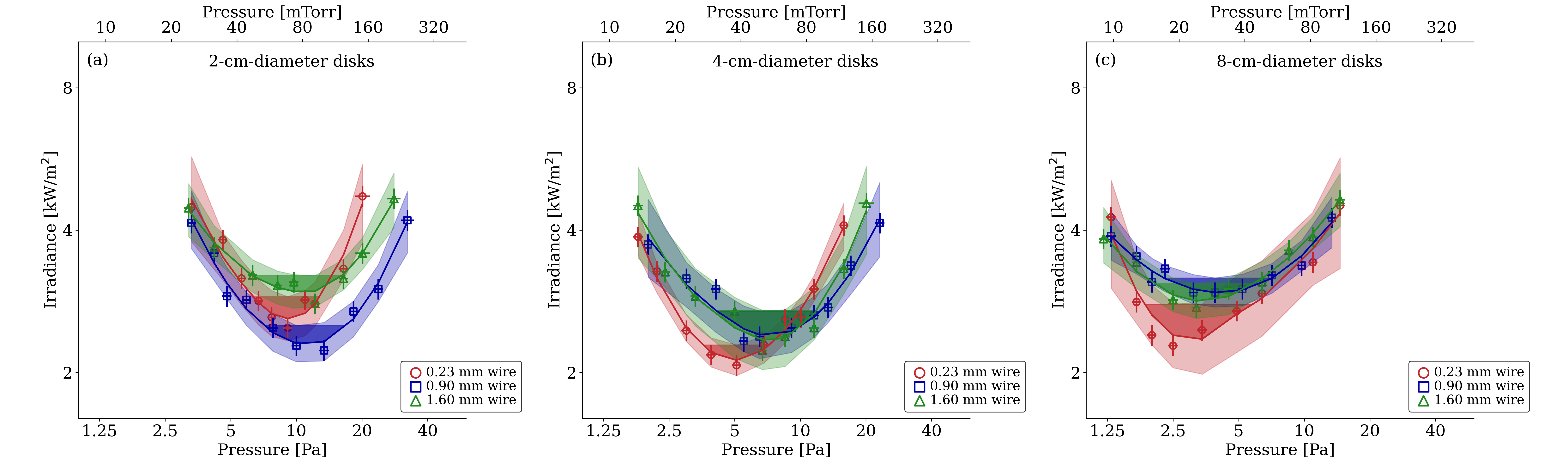}
\caption{Same as Fig. 2 for cane-shaped steel sticks with changing wire diameters. The optimal pressure can be seen to be dependent on the microflyer diameter and independent on the wire diameter. Note that the irradiance axis’s range is adjusted from Fig. 2 and 3 to show more clear irradiance-pressure patterns.}
\end{figure*}

By comparing the results for the first-generation 0.23-mm-diameter steel mesh launchpad (Fig. 3) with the second-generation 0.23-mm-diameter J-shaped stick launchpad (Fig. 4), it can be seen that the minimum light irradiances necessary for lift-off were similar. This indicates that an 88\%-open steel mesh approaches the performance of the minimal launchpad. This is corroborated by the fact that we did not detect any large differences between 73\%-open and 85\%-open meshes (both high open area percentages) in our previous experiments \cite{controlled-levitation}. However, for the thickest 1.6-mm-diameter wires, the minimum required irradiances can be seen to be up to 6 times lower for meshes than for second-generation launchpads. Since the light irradiance scales approximately linearly with the photophoretic lift force in the transition regime \cite{controlled-levitation, semi-emipirical}, for a specific microflyer with a given weight, this sixfold decrease in irradiance is approximately equivalent to a sixfold enhancement of the photophoretic lift force due to the ground effect.  

The basic mechanism by which the wire mesh increases the photophoretic lift force on microflyers, e.g., the ground effect, can be described as follows. As shown in Fig. S2(a), for closely located horizontal surfaces wherein the mean free path is larger than the inter-surface spacing, air molecules can more frequently bounce back and forth between the two surfaces, imparting recoil forces that are larger than those experienced by a microflyer in mid-air. Right before the takeoff, the disk rests on the launchpad and the wire diameter determines the typical distance that the molecules travel when bouncing back and forth between the levitating disk and the launchpad, which sets the characteristic length scale and explains why the optimal pressure depends on the wire diameter. Similarly, the magnitude of the ground effect also gradually decreases with increasing open area and eventually vanishes as the mesh open area approaches 100\%. Therefore, as the underlying surface becomes increasingly sparse, the dependence on the open area percentage and the wire diameter becomes insignificant, and the optimal pressure instead begins to depend on the size of the levitating disk. Note that rigorous modeling of the observed ground effect requires computer-intensive numerical modeling in the transition regime \cite{ground-effect, computational-study, aerodynamic-ground}, which is beyond the scope of the current experimentally-focused manuscript.

Even with ultra-sparse J-shaped sticks that themselves yield insignificant ground effects, other underlying surfaces may become important in producing a floor-associated ground effect due to their vertical proximity to the microflyers \cite{numeric-investigation, in-ground}. After the launchpad, the second-closest horizontal surface for any sample is the vacuum chamber’s floor. To investigate the associated ground effect, we prepared J-shaped sticks with 2, 3, and 4 cm lengths and changed the vertical distance between the sample and vacuum chamber’s bottom surface while maintaining the vertical LED-sample distance (by moving the chamber itself) to keep the light irradiance on the microflyers constant. As shown in Fig. 5, the smallest 2-cm-diameter disks have very similar optimal pressures and irradiances regardless of the floor-to-microflyer distance, while 4 and 8-cm-diameter microflyers both have lower optimal pressures and greater minimum light irradiances as the distance grows. These observations indicate that the floor-associated ground effect due to the chamber’s bottom gradually decreases as the distance from the disk to the chamber bottom increases, and moreover that it becomes insignificant when this distance exceeds the disk diameter. Similar to large aerial vehicles like rotorcraft and hovercraft \cite{wing-in, design-of, aerodynamic-investigation, ground-effect, mitigating-ground}, this type of ground effect results from the underlying surface deflecting the airflow around the microflyer downward and thus creating an area of higher pressure (i.e., an air cushion) under the microflyer, as sketched in Fig. S2(b, c).

\begin{figure*}[hbt!]
\centering
\includegraphics[width=\textwidth]{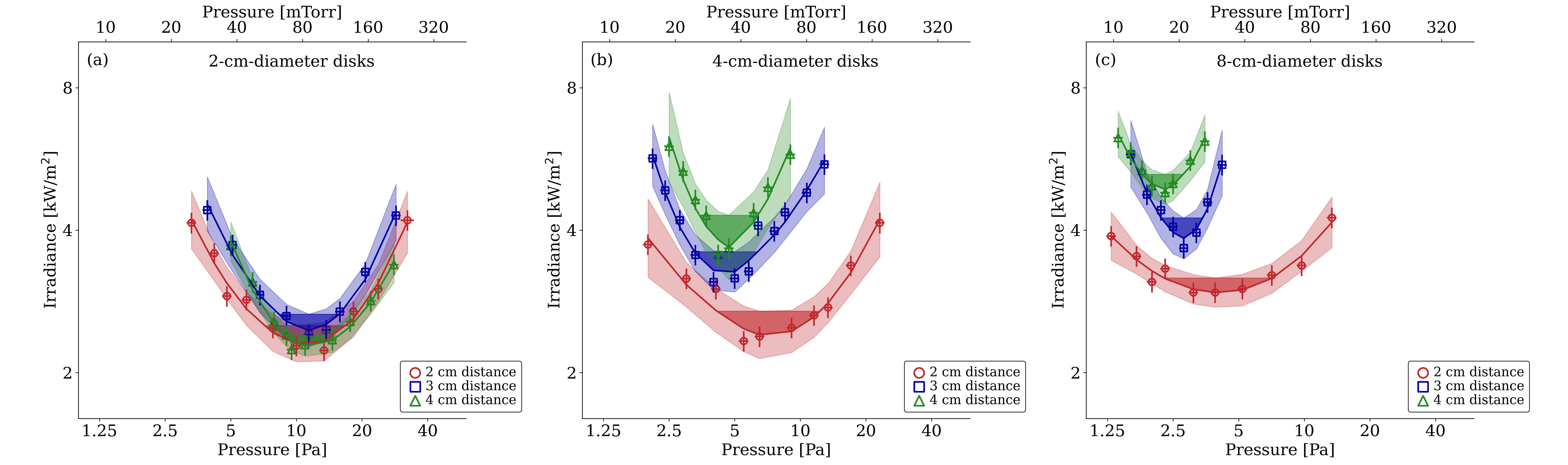}
\caption{Same as Fig. 2 for cane-shaped steel sticks with a fixed wire diameter (i.e., 0.9 mm) but changing distances to the chamber's floor. The optimal pressure can be seen to be dependent on the floor-to-disk distance when that distance is smaller than the disk diameter. Note that the irradiance axis’s range is adjusted from Fig. 2 and 3 to show more clear irradiance-pressure patterns.}
\end{figure*}

Thus, we can suggest the following guidelines when testing microflyers in vacuum chambers. First, the launchpad-associated ground effect should be minimized by making the launchpad so sparse that the optimal pressure no longer depends on the parameters of the launchpad and instead scales with the size of the microflyer. Second, the distance between the vacuum chamber floor and the microflyer must be increased until pressure-irradiance curves cease to show a vertical distance dependence. Typically, the distance needs to be larger than the largest dimension of the levitating structure to minimize the floor-associated ground effect. We finally note that the ground effect in the transition regime, typical for photophoretic experiments, will exhibit aerodynamic features of both the free molecular and continuum regimes (Fig. S2).

\section{\label{sec:level1}Conclusion}
In summary, we demonstrated that photophoretic microflyers tested in relatively small vacuum chambers can experience large ground effects associated with both the supporting launchpad structures and chamber bottom, which may lead to greatly exaggerated expectations of the photophoretic lift force in mid-air applications. We characterized steel mesh launchpads with different wire diameters and spacings, where we observed stronger ground effects for denser meshes. Then, we developed a minimal launchpad of J-shaped steel sticks that was nearly 100\% open as to make the launchpad-associated ground effect’s impact insignificant, as indicated by the irradiance-pressure graphs’ independence of the wire diameter. Furthermore, we varied the distance between the microflyer disks and the vacuum chamber’s bottom surface, concluding that the distance between the microflyer and any other underlying surface should be at least as large as the diameter of the disk to minimize the floor-associated ground effect. Minimizing the ground effect in laboratory tests of photophoretic microflyers is important for developing realistic expectations for the payloads of photophoretic UAVs in Earth’s mesosphere or the atmosphere of Mars.

\begin{acknowledgments}
This work is partly supported by the NSF under CBET-1845933 and the School of Engineering and Applied Science at the University of Pennsylvania.
This work was carried out in part at the Singh Center for Nanotechnology, which is supported by the NSF National Nanotechnology Coordinated Infrastructure Program under grant NNCI-2025608.
\end{acknowledgments}

\section*{Data Availability Statement}
The data that support the findings of this study are available from the corresponding author upon reasonable request.

\section*{Conflict of Interest}
The authors have no conflicts to disclose.

\section*{Author Contributions}
\textbf{Zhipeng Lu}: conceptualization (equal); methodology (equal); resources (lead); investigation (lead); data curation (lead); formal analysis (equal); software (lead); writing - original draft (lead). \textbf{Miranda Stern}: resources (lead); investigation (supporting). \textbf{Jinqiao Li}: resources (supporting); software (supporting). \textbf{David Candia}: resources (supporting). \textbf{Lorenzo Yao-Bate}: investigation (supporting). \textbf{Thomas J. Celenza}: resources (supporting); investigation (supporting). \textbf{Mohsen Azadi}: conceptualization (supporting); methodology (supporting); software (supporting). \textbf{Matthew F. Campbell}: writing - review and editing (supporting). \textbf{Igor Bargatin}: conceptualization (equal); formal analysis (equal); writing - review and editing (lead).\\

\appendix*
\section{Materials and Methods}
\subsection{\label{app:subsec}Procedure for microflyer fabrication}
We started with a 0.5-$\mu$m-thick mylar sheet (Dupont) whose area density was $\sim$0.7 g/m$^2$ as measured on an analytical balance (A\&D HR-202). We wrapped the film around a 525-$\mu$m-thick silicon wafer and spin-coated a solution of 0.2\% (weight) water-based single-wall CNT (1-2 nm diameter and 5-30 $\mu$m length, NanoAmor) on the top surface at 300 revolutions per minute for 10 seconds. We baked the resulting bilayer structure on a hotplate at 90 °C for 10 minutes. By weighing the samples, we determined the areal density to be 0.9–1.3 g/m$^2$ after this step. We then flipped the mylar-CNT film and deposited a layer of 100-nm-thick alumina via atomic layer deposition at 140 °C using water and Al$_2$(CH$_3$)$_6$ precursors (Cambridge Nanotech S2000 ALD). Last, we used laser micromachining (IPG IX280-DXF) to cut circular disks with 2, 4, and 8 cm diameters. The areal density of final alumina-mylar-CNT disks was typically 1.2–1.6 g/m$^2$.

\subsection{\label{app:subsec}Procedure for vacuum chamber testing}
We used a customized 10-liter cylindrical vacuum chamber with an acrylic body and steel flanges as pictured in Fig. S1(a, b). A two-stage (roughing-turbo) vacuum pump (Pfeiffer HiCube 80 Eco Turbo Pumping Station) allowed us to reach chamber pressures between 0.8 and 200 Pa, as measured by a vacuum gauge sensor (InstruTech, Inc., CVG101GF). In experiments using metal meshes, the meshes were electrically grounded to minimize electrostatic forces. In experiments using J-shaped steel sticks, we added a 1-mm-thick polyethylene terephthalate film coated by indium tin oxide (ITO-PET film from Adafruit) underneath the ring holder to provide electrical grounding. This optically transparent and electrically conductive film formed an effective and convenient electrical grounding circuit with $\sim$85\% optical transparency (we accounted for this partial absorption when calculating the actual irradiance on the microflyers). We set up an 8-LED (LOHAS LH-XP-100W-6000K) array to create a symmetric, uniform, and sufficiently intense light source that could be tuned continuously through a power supply (Teyleten Robot Non-Isolated Step-Up Module). Finally, we note that we applied a thin layer of silver paste (Arctic Silver 5 Polysynthetic Thermal Compound) between the LEDs and aluminum base plate to enhance the heat dissipation from the LED array. 

\subsection{\label{app:subsec}Procedure for light irradiance characterization}
The whole LED array could safely provide light irradiances of up to 7 kW/m$^2$ (absent any shadowing from the launchpad), as measured using optical power and energy sensors. The methodology is described as follows: (1) using collimated light from the LED array to establish an equivalence of signals from a fully open photodiode (Vishay Semiconductors Silicon PIN Photodiodes Osram BPW34) and a partially open optical sensor (Thorlabs, Inc., S305C and PM100USB); (2) finding the percentage of light received by the optical sensor by repeating Step (1) but using the original light from the LED array; (3) measuring the light irradiance with the percentage from Step (2) considered. Note that the shadowing effect of every launchpad was characterized and both photodiode and optical sensor were pre-calibrated.

\subsection{\label{app:subsec}Experimental results of previous publications}
Cortes et al. \cite{photophoretic-levitation} reported photophoretic levitations of nanocardboard rectangular plates on (1) a micropatterned 0\%-open glass substrate, and (2) an 84\%-open wire mesh. The plates were 6 mm by 13 mm in dimension and 0.1 mg in weight. The levitation height was 10 mm. We argue that the launchpad and floor-associated ground effects from the glass launchpad were potentially large, while those from the 84\%-open wire-mesh launchpad were much less significant.

Azadi et al. \cite{controlled-levitation} reported photophoretic levitations of circular disks on (1) a 74\%-open wire mesh, and (2) an 85\%-open wire mesh. The disks were 6 mm in diameter and 0.03 mg in weight. The levitation height was about 5 mm. We argue that the launchpad and floor-associated ground effects from both meshes were not large and reasonably showed slight differences in experimental results. 

\setcounter{figure}{0}
\renewcommand{\figurename}{Fig.}
\renewcommand{\thefigure}{S\arabic{figure}}
\begin{figure}[hbt!]
\centering
\includegraphics[width=\columnwidth,trim={0 8cm 11cm 0},clip]{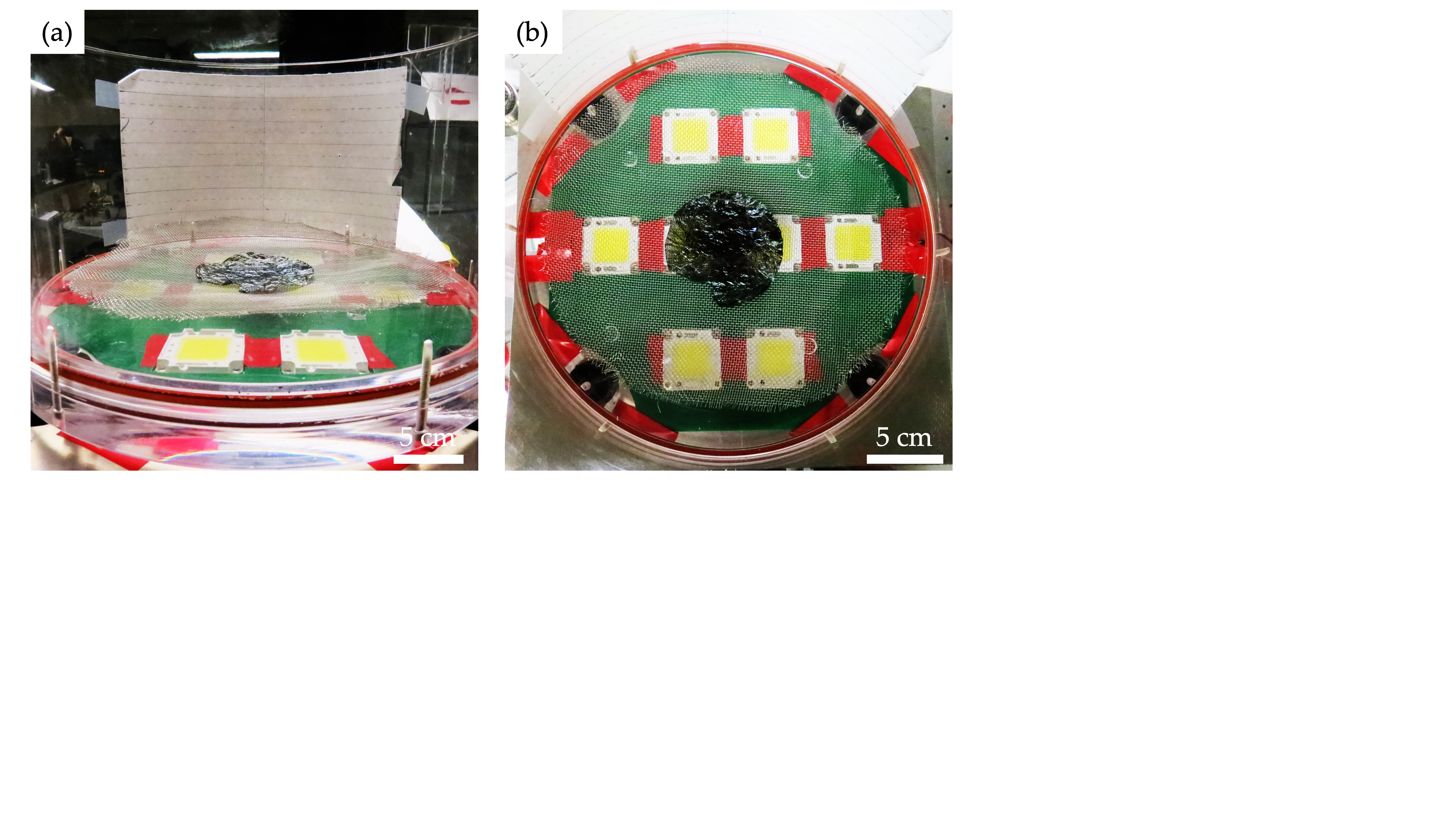}
\caption{Photographs of experimental setups consisting of an acrylic vacuum chamber, an 8-cm-diameter CNT-mylar-alumina microflyer, a piece of 73\%-open steel mesh, and eight LEDs below the chamber. (a) Side view. (b) Top-down view.}
\end{figure}

\begin{figure}[hbt!]
\centering
\includegraphics[width=\columnwidth,trim={0 12cm 7cm 0},clip]{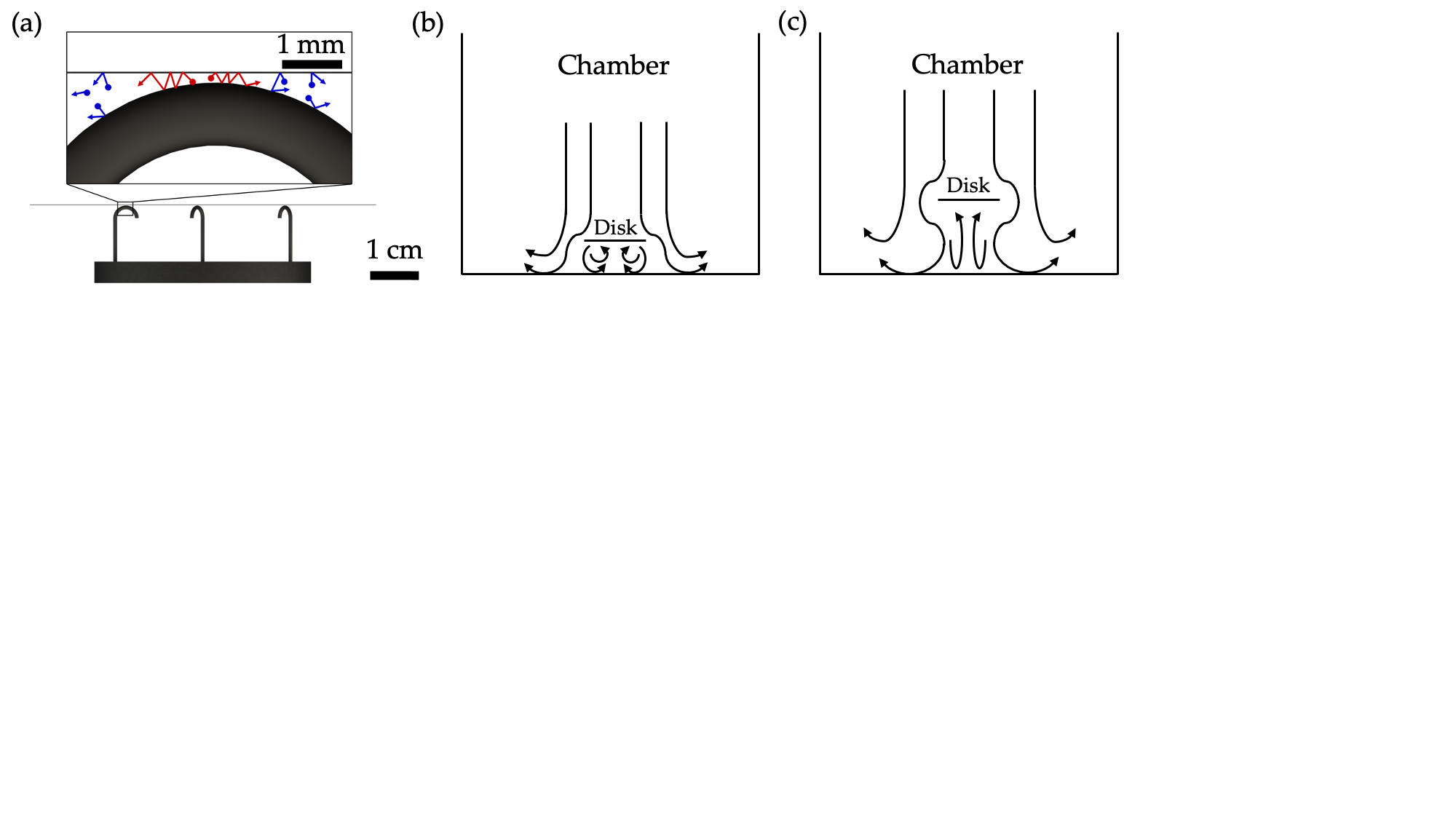}
\caption{Simplified gas-dynamic features between the microflyer and neighboring surfaces. (a) Cross-sectional schematic diagrams of the molecule trajectories during the levitation on J-shaped steel sticks. (b, c) Cross-sectional schematic diagram of the air flow during the levitation when the floor-to-disk distance is (b) smaller than the disk’s diameter and (c) greater than the disk’s diameter. Note that in (a), red arrows represent molecules bouncing back and forth between the launchpad wire and the levitating disk, leading to the ground effect, while blue arrows represent other trajectories of air molecules.}
\end{figure}

\nocite{*}
\bibliography{paper}

\end{document}